\documentclass[12p]{article}
\usepackage{latexsym}
\usepackage{color}
\usepackage{amsmath}
\usepackage{epsfig}
\usepackage{hyperref}
 \usepackage{dcolumn}
\newcommand{\ortala}[1]{\begin{center}#1\end{center}}

\newcommand{\ket}[1]{\left|#1\right\rangle}

\newcommand{\sand}[3]{\left\langle #1\left|#2\right|#3\right\rangle}
\newcommand{\sandd}[1]{\left\langle #1\right\rangle}

\newcommand{\integ}[3]{{{\underset{#1 }{\overset{#2}{\displaystyle\int}}}#3}}

\newcommand{\summ}[3]{{{\underset{#1 }{\overset{#2}{\displaystyle\sum}}}#3}}

\newcommand{\re}[1]{(\ref{#1})}

\newcommand{\eq}[2]{\begin{equation}\label{#1}  #2\end{equation}}

\newcommand{\paran}[1]{\left(#1\right)}

\newcommand{\sch}[1]{Schrodinger}
\newcommand{\ktur}[2]{\frac{\partial #1}{\partial #2}}

\usepackage{amssymb}
\usepackage{latexsym}
\begin{document}

\ortala{\large\textbf{Magnetocaloric effect in bilayer material: Spin-S Ising model}}

\ortala{\textbf{\"Umit Ak\i nc\i\footnote{\textbf{umit.akinci@deu.edu.tr}}, Yusuf Y\"uksel }}

\ortala{\textit{Department of Physics, Dokuz Eyl\"ul University,
TR-35160 Izmir, Turkey}}

\section{Abstract}

The isothermal magnetic entropy change has been obtained for the bilayer system that consists of $S_u$ and $S_l$ valued spins in each layer. The relation between the IMEC and the spin value as well as the value of exchange interaction between two layers have been obtained. Recently experimentally given double peaks behavior for manganite bilayers has been demonstrated theoretically, and explanation is given. On the other hand, the conditions that give rise to transition from this double peak behavior to the single peak behavior has been obtained and discussed.

\section{Introduction}\label{introduction}

Magnetocaloric effect (MCE) is one of the promising applications of magnetism in cooling technology. 
In analogous to the traditional compression-expansion thermal-mechanical cycle, which is used for 
cooling applications, MCE is an alternative to this technology, which promises clean energy.  
It was first observed in Iron \cite{ref1}, and after this discovery,  explanations were introduced theoretically
\cite{ref2,ref3}. MCE is simply realized in a compression-expansion like a thermodynamical cycle. The adiabaticity
of the process forces the total entropy of the material to remain constant. But in different steps of the cycle, by applying a magnetic field, the magnetic part of the entropy changes. This change is balanced by the lattice contribution to the total entropy, then the temperature of the material changes.
The efficiency of the MCE can be determined by refrigerant capacity (RC). This quantity is defined via
isothermal magnetic entropy change (IMEC) of the material. 
In this context, a good candidate for magnetic refrigerant material should have a
 large magnitude of IMEC spreaded in a broader temperature span. 
These fundamental parts of the MCE are today well defined 
\cite{ref4}.  

On the other hand, nanomagnetic systems constitute a growing research area due to the vast technological applications and
enhance our understanding related to the nanoworld, theoretically. Confinement of the material changes its magnetic properties (as well as other properties such as electronic properties) drastically.  These changes in the
magnetic properties may promise that nanosystems can be good candidates for magnetic refrigerant material. 
For instance, it has been shown for thin films and nanoparticles, the behavior spreading to a broader temperature span in IMEC can yield slightly increased refrigerant capacity \cite{ref5}. 
The enhancement of the magnetocaloric performance of the bilayer system was discussed in \cite{ref6}. The experimental double peak behavior in IMEC for bilayers have been obtained that work. For this possible enhancement of MCE in 
magnetic bilayers, experimental studies related to these systems were performed. 
Magnetocaloric properties of bilayer perovskite manganite $La_{1.38} Sr_{1.62} Mn_2 O_7$ have been determined
experimentally. The system has a Curie temperature around 114 K, which is the temperature of the second-order transition from the ferromagnetic phase to the paramagnetic phase \cite{ref7}. 
Magnetocaloric properties of $Fe_{48} Rh_{52}$ - $PbZr_{0.53} Ti_{0.47} O_3$ bilayer have been investigated
experimentally and large magnetoelectric ordering found around the antiferromagnetic
phase transition temperature of $315 K$ \cite{ref8}. This transition temperature promises that this material may be applicable to everyday applications.
Co and Fe based monolayer and their corresponding bilayer forms have been prepared, and values of the 
IMEC, RC, and adiabatic temperature change have been determined by applying the magnetic field up to $1T$ \cite{ref9}.
The double-layered perovskite manganite structure $Pr_{1.75} Sr_{1.25} Mn_2 O_7$ was 
synthesized and MCE properties have been determined. Second-order phase transition of this material was obtained as $305K$.  
IMEC and adiabatic temperature change have been determined by applying the magnetic field up to $70 kOe$ \cite{ref10}.

As the experimentalists' interest in the topic grows, on the theoretical side there is no complete understanding of the MCE in the bilayer. But, on the theoretical side, there are numerous works related to the magnetic properties of the bilayer systems. For instance, 
spin-1/2 spin -1 Ising bilayer system has been worked within the effective field approximation \cite{ref11}. 
A similar system, which is a double layer superlattice consists of  spin-1/2 spin -1 portions have been worked within the Monte Carlo (MC) simulations \cite{ref12}. The effect of the indirect exchange interactions (modeled by nonmagnetic middle
layer)
on the magnetic properties of the spin-1/2 spin -1 Ising bilayer has been determined by effective field theory \cite{ref13}. 
The magnetic and hysteresis behaviors of higher spin bilayer have also been determined by MC simulations and mean-field 
approximations such as mixed spin-1 and spin-3/2 Ising bilayer \cite{ref14} and  Spin-7/2 and Spin-5/2 Ising 
bilayer \cite{ref15}.

Besides, the magnetic properties of the spin-1/2 Ising-Heisenberg bilayer have been determined by pair approximation. 
The double peak behavior in specific heat and IMEC have been observed \cite{ref16}. The same  double peak behavior in IMEC has been determined for  spin-1 and spin-2 Heisenberg superlattice, 
theoretically by using Green's function method \cite{ref17}.

As shown in this short literature, up to now, there is no general conclusion about the bilayer system, which consists of arbitrary spin value. Then, the aim of this work is to determine the dependency of the MCE efficiency of the magnetic bilayer 
on the spin value and the value of exchange interaction. 
For this aim, the paper is organized as follows: In Sec. 
\ref{formulation} we
briefly present the model and formulation. The results and
discussions are presented in Sec. \ref{results}, and finally Sec.
\ref{conclusion} contains our conclusions.

\section{Model and Formulation}\label{formulation}


\begin{figure}[h]\begin{center}
\epsfig{file=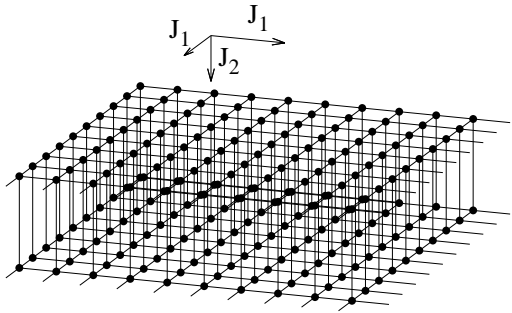, width=10.0cm}\end{center}
\caption{Schematic representation of the bilayer system} \label{sek1}
\end{figure}

We can see the schematic representation of the bilayer geometry in Fig. \ref{sek1}.  As seen in 
Fig. \ref{sek1}, one layer of the system consists of a regular lattice with coordination number $z$. Fig. \ref{sek1} depicts a
square lattice, i.e. $z=4$. Let the upper layer spins have value $S_u$ and the lower layer have $S_l$. We can write 
the Hamiltonian of this system as
\eq{denk1}{
\mathcal{H}=-J_1\summ{\alpha=u,l}{}{\summ{<i,j>}{}{S_i^\alpha S_j^\alpha}}
-J_2\summ{i}{}{S_i^uS_i^l}
-H\summ{\alpha=u,l}{}{\summ{i}{}{S_i^\alpha}}
,}
where $S_i^u,S_i^l$ denote the $z$ component of the Pauli spin 
operator at a site $i$ at layers denoted by $u$ (upper) and $l$ (lower), respectively. 
$J_1$ and $J_2$ are the intralayer/interlayer exchange interaction between the nearest neighbor spins, respectively. 
$H$ is the longitudinal magnetic field. The first 
summation in Eq. \re{denk1} is taken over the nearest neighbor sites in a layer, the second summation is 
taken over the nearest neighbor sites which belong to different layers, 
and the third summation is taken over all the lattice sites. 

One can construct several different finite clusters and derive the formulation related to that cluster. The effect of the size of the cluster on the accuracy of the solution can be found in Ref. \cite{ref18}. If we choose one spin 
cluster we obtain the traditional solution of mean-field or effective-field approximation, according to the handling of local fields.
In order to get more accurate results within
the local field approximations, we can choose 4-spin cluster: two spins from the upper plane and two nearest neighbors of them from lower plane. Note that, by repeating to this chosen finite cluster we can construct a bilayer system completely. 

The Hamiltonian of this finite cluster can be written from Eq. \re {denk1} as 
\eq{denk2}{
\mathcal{H}^{(4)}=-J_1\paran{S_1^u S_2^u+S_1^l S_2^l}
-J_2\paran{S_1^u S_1^l+S_2^u S_2^l}
}
$$
-H\paran{S_1^u + S_2^u+S_1^l+  S_2^l}
-\paran{h_1^u S_1^u+h_2^u S_2^u+h_1^l S_1^l+h_2^l S_2^l}, 
$$
where  $h_i^\alpha$ is the local field that represents the interaction of the $i^{th}$
spin in a layer $\alpha=u,l$, with nearest-neighbor spins 
belonging to outside of the cluster.

The thermal average of  a quantity $\Omega$ can be calculated via 
the exact generalized
Callen-Suzuki identity \cite{ref19}
\eq{denk3}{
\sandd{\Omega}=\sandd{\frac{Tr_4 \Omega \exp{\paran{-\beta 
\mathcal{H}^{(4)}}}}{Tr_4 \exp{\paran{-\beta \mathcal{H}^{(4)}}}}}.
} In Eq.  \re{denk3}  $Tr_4$ stands for the partial trace over all the lattice 
sites which belong to the chosen cluster and $\beta=1/(kT)$, where $k$ is the 
Boltzmann constant and $T$ is the temperature. Note that, since the model is Ising model, matrix 
representation of $\mathcal{H}^{(4)}$ is diagonal, then exponential can be handled easily. For the aim of finding 
the matrix representation of $\mathcal{H}^{(4)}$, let us construct 4-spin bases as  
 $\{\phi_i\}=\ket{s_1^u s_2^u  s_1^l s_2^l }$, where 
$s_k^\alpha $ is just one spin eigenvalues of the operator $S_k^\alpha$  
($k=1,2$). Remember that,  $S_k^\alpha$ is the z component of spin-$S_\alpha$ operator, where $\alpha =u,l$.

The well known effect of the spin operator on a base can be written as 
\eq{denk4}{
S_k^\alpha\ket{s_1^u s_2^u s_1^l s_2^l }=s_k^\alpha
\ket{s_1^u s_2^u  s_1^l s_2^l } ,
} where $k=1,2$ and $\alpha=u,l$.
Note that, since the system consist of spin-$S_u$ in upper layer and spin-$S_s$ particles in lower layer, 
number of bases equals to $\paran{2S_u+1}^2\paran{2S_l+1}^2$.

As we said before only non-zero elements are diagonal elements, and they can be obtained 
by applying the operator given in Eq. \re{denk2} to bases, according to Eq. \re{denk4}. 
They are given by,

\eq{denk5}{
\Phi\paran{s_1^u,s_2^u,s_1^l,s_2^l}=\sand{\phi_i}{\mathcal{H}^{(4)}}{\phi_i}=-J_1\paran{s_1^u s_2^u+s_1^l s_2^l}
-J_2\paran{s_1^u s_1^l+s_2^u s_2^l}
}
$$
-H\paran{s_1^u + s_2^u+s_1^l+  s_2^l}
-\paran{h_1^u s_1^u+h_2^u s_2^u+h_1^l s_1^l+h_2^l s_2^l}. 
$$

Then we can write Eq.  \re{denk3} as 

\eq{denk6}{
\sandd{S_k^\alpha }=\sandd{\frac{\summ{\{s_1^u,s_2^u,s_1^l,s_2^l\}}{}{} s_k^\alpha  \exp{\paran{-\beta 
\Phi }}}{\summ{\{s_1^u,s_2^u,s_1^l,s_2^l\}}{}{} \exp{\paran{-\beta 
\Phi }}}}, k=1,2; \alpha=u,l. 
} The summations are taken over all possible configurations of $(s_1^u,s_2^u,s_1^l,s_2^l)$. 

We assume that, layers are translationally invariant, i.e. $\sandd{S_1^u}=\sandd{S_2^u}=m_u$ and 
$\sandd{S_1^l}=\sandd{S_2^l}=m_l$. The layer magnetizations and total magnetization can be defined by

\eq{denk7}{
m_\alpha=\frac{1}{2}\paran{\sandd{S_1^\alpha}+\sandd{S_2^\alpha}}, \quad 
\quad m=\frac{1}{2}\paran{m_u+m_l}.
}

Since the local fields, $h_i^\alpha$ represent the spin interactions of $S_i^\alpha$ with outside spins,
they can be written in a mean-field level as

\eq{denk8}{
h_i^\alpha=(z-1)J_1 m_\alpha. 
} By using this approximation, nonlinear equations given in Eq. \re{denk6} can be written as 
\eq{denk9}{
\sandd{S_k^\alpha }=\frac{\summ{\{s_1^u,s_2^u,s_1^l,s_2^l\}}{}{} s_k^\alpha  \exp{\paran{-\beta 
\Phi }}}{\summ{\{s_1^u,s_2^u,s_1^l,s_2^l\}}{}{} \exp{\paran{-\beta 
\Phi }}}, k=1,2; \alpha=u,l. 
} 
The number of four nonlinear equations seen in Eq. \re{denk9} can be solved numerically. By using these solutions in 
Eq. \re{denk7}, layer and total magnetizations can be obtained.

In  order to determine the magnetocaloric properties of the system, we calculate the 
IMEC when the maximum applied  longitudinal field is $H_{max}$. The definition of IMEC is given by
\eq{denk10}{
\Delta S_M=\integ{0}{H_{max}}{}\paran{\ktur{m}{T}}_H dH.
} 



\section{Results and Discussion}\label{results}


Let us focus our attention on the magnetocaloric properties of the bilayer system, which is consists of interacted 
spin-$S_u$ and spin-$S_l$ layers. 
By choosing a unit of energy as $J_1$, we can work with scaled  quantities as
$$
r=\frac{J_{2}}{J_1}, \quad h=\frac{H}{J}, \quad t=\frac{k_BT}{J}.  
$$

Note that, $h_{max}= 1.0$ throughout the calculations. The system consists of an interacting two-layers, which have completely the same geometry. Then interchanging spin values 
$S_u$ and $S_l$ do not change the results. In other words, calculations for $(S_u,S_l)$ pair and $(S_l,S_u)$ pair
exactly yield the same results. Since the magnetization behavior and critical properties of the bilayer system are well studied, let us start our discussion by variation of the IMEC for different bilayer systems.

\begin{figure}[h]\begin{center}
\epsfig{file=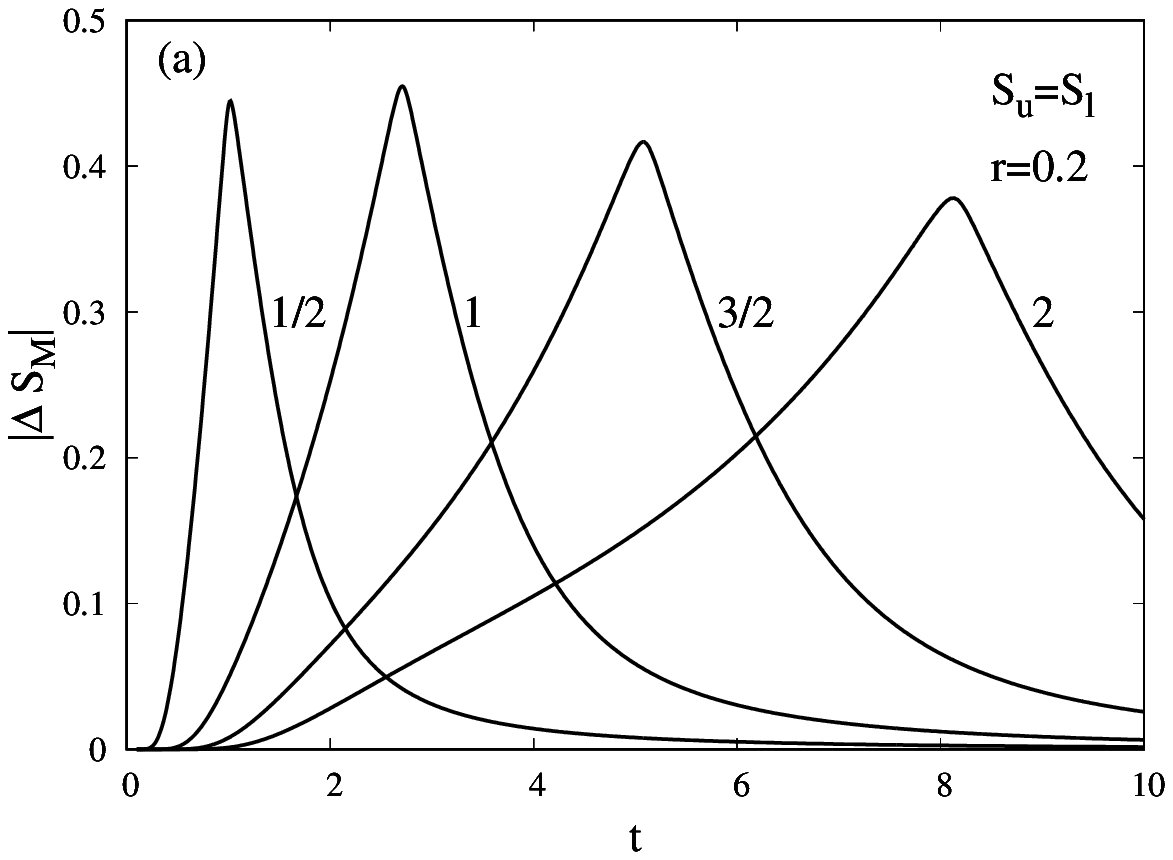, width=6.0cm}
\epsfig{file=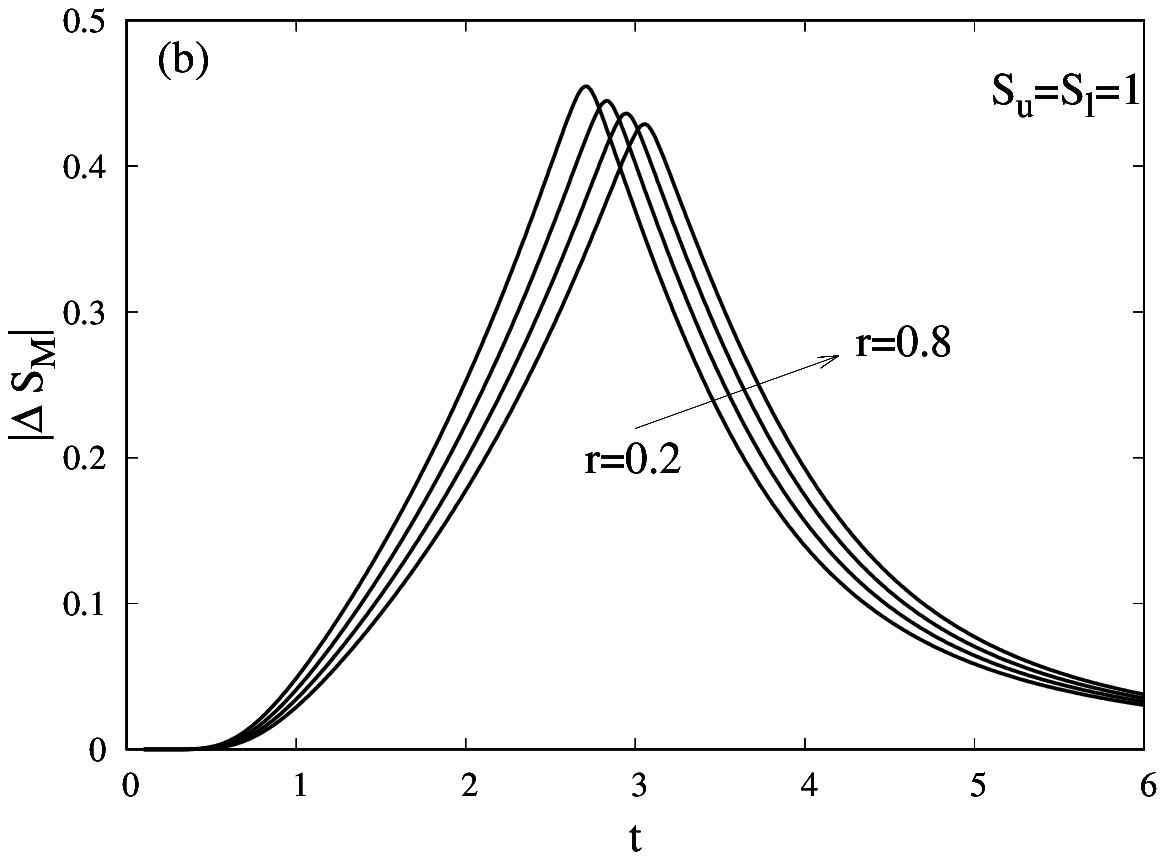, width=6.0cm}
\end{center}
\caption{The variation of IMEC with the temperature for selected values of (a) $r=0.2$ and
$S_u=S_l=1/2,1,3/2,2$, (b) $S_u=S_l=1$ and $r=0.2,0.4,0.6,0.8$ 
for the bilayer system.} \label{sek2}
\end{figure}

\begin{figure}[h]\begin{center}
\epsfig{file=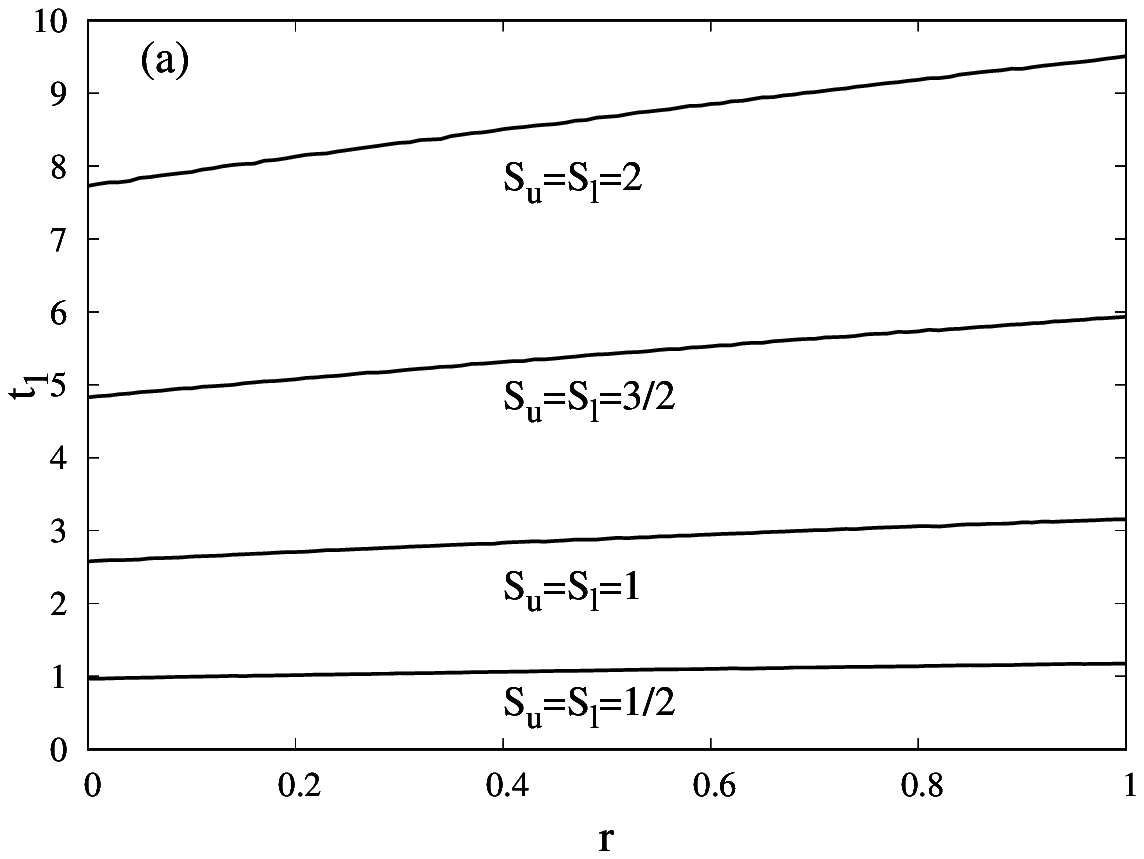, width=6.0cm}
\epsfig{file=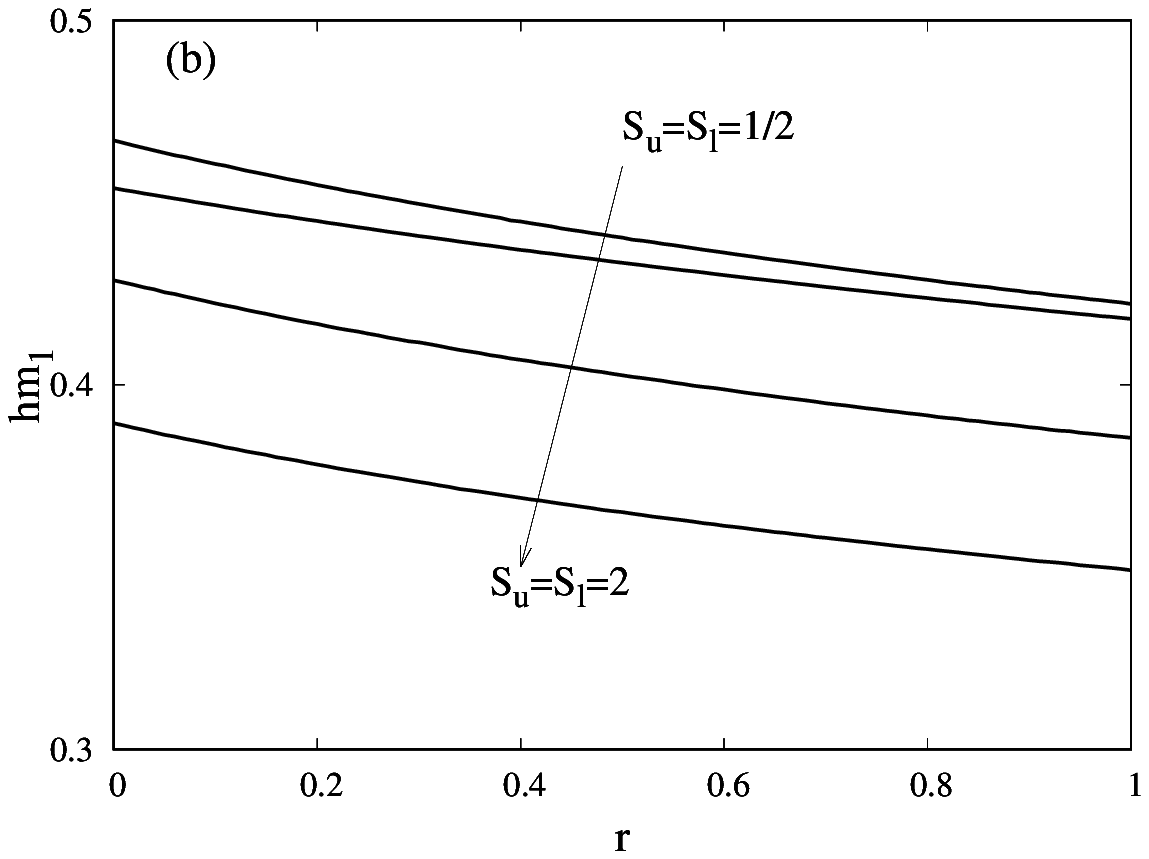, width=6.0cm}
\end{center}
\caption{The variation of the (a) location of the peak and (b) height of the peak in IMEC versus temperature curves for some selected values of the spins. These values have been provided on the Figure.} \label{sek3}
\end{figure}

We start with the case $S_u=S_l$, i.e., bilayer, which consists of the same valued spins. We can see the effect of the spin value on the variation of the IMEC with the temperature in Fig. \ref{sek2} (a) and the effect of the exchange interaction between two layers of the system in Fig. \ref{sek2} (b). As seen in Fig. \ref{sek2} (a), rising spin value shifts the peaks in IMEC curves towards the right of the $(|\Delta S|_M,t)$  plane. This is expected because it is a well-known fact that the peak is located near the critical temperature of the system. Larger spin value means the larger critical temperature for the system. At the same time, the peak values get slightly lower, while the value of the spin increases. This is also consistent with the literature \cite{ref20}. The same behavior is valid for rising exchange interaction between the layers, for arbitrarily chosen spin values (see Fig. \ref{sek2} (b)). General trends related to these behaviors can be seen in Figs. \ref{sek3} (a) and (b). Rising exchange interaction value between the layers increases the location ($t_1$) of the peak in IMEC and decreases the height of the peak ($hm_1$). Note that $hm$ in Fig. \ref{sek3} (b) stands for the height of the peak (maximum of the IMEC), and indices of $t$ and $hm$ is necessary since there are two peaks in the IMEC curve for the case $S_u\neq S_l$, as explained below.

\begin{figure}[h]\begin{center}
\epsfig{file=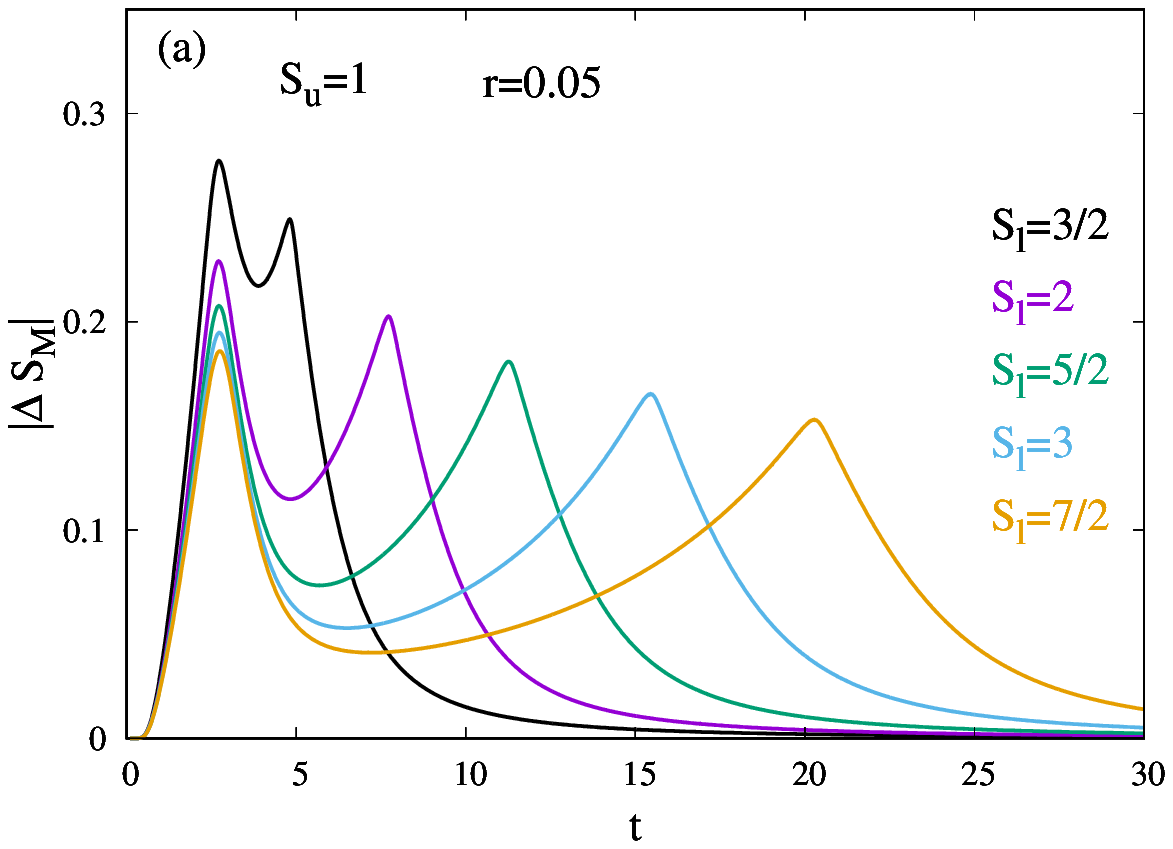, width=6.0cm}
\epsfig{file=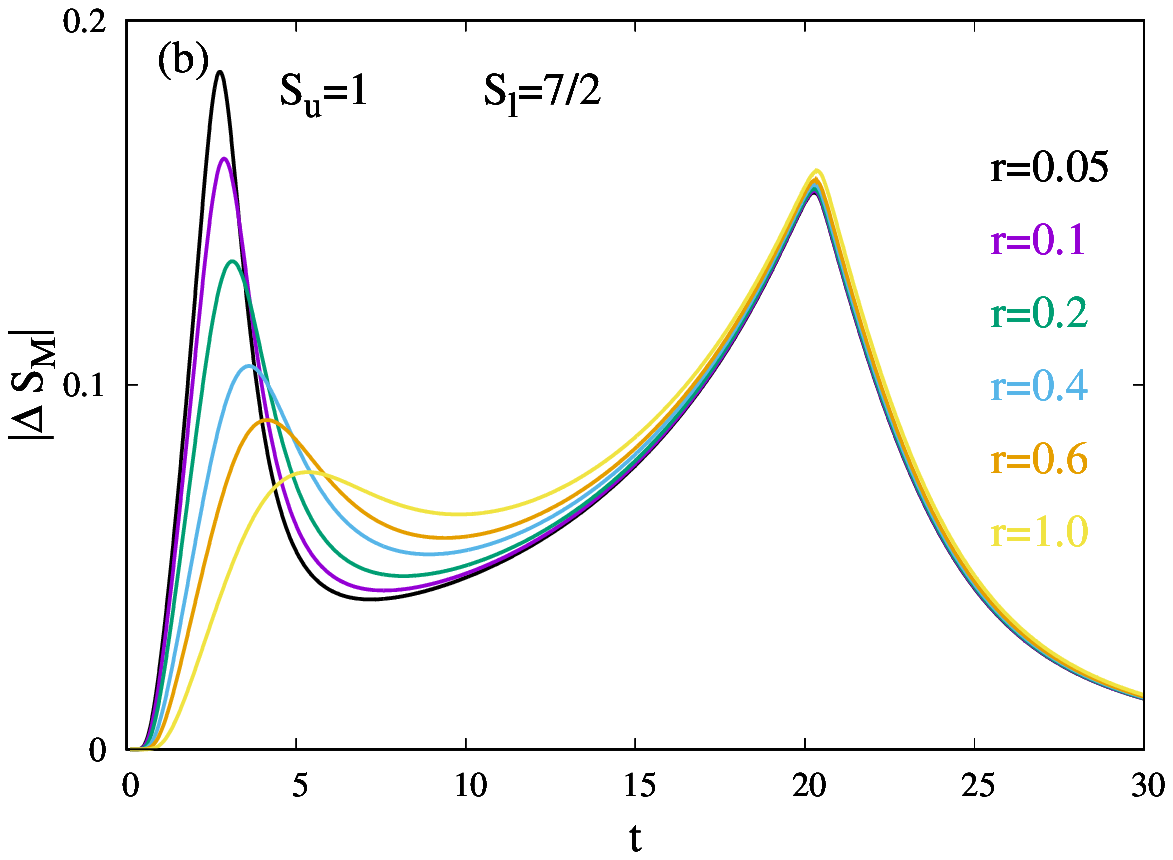, width=6.0cm}
\end{center}
\caption{The variation of IMEC with the temperature for selected values of (a) $S_u=1,r=0.05$ and
$S_l=3/2,2,5/2,3,7/2$, (b) $S_u=1,S_l=7/2$ and $r=0.05,0.1,0.2,0.4,0.6,1.0$ 
for the bilayer system.} \label{sek4}
\end{figure}

Our investigation is continues for the bilayer system that is constituted by layers that have different valued spins. 
By convention, we choose $(S_u,S_l)$ pairs such that, $S_u\leq S_l$. 
In Fig. \ref{sek4} (a) we depict the variation of IMEC with the temperature for weakly interacted two-layer system 
$(r=0.05)$  with spin values $S_u=1$ and $S_l=3/2,2,5/2,3,7/2$. As seen in Fig. \ref{sek4} (a), all bilayers display two peaks in variation of the IMEC  with the temperature. This double peak behavior is already obtained experimentally
\cite{ref6} and theoretically \cite{ref16}, \cite{ref17}.  
This  behavior is due to the different 
critical temperatures of the single layers when they are independent of each other.
Weak interaction allows two-layers to act almost independently of each other. When the $S_l$ gets higher values, the second peak of the system moves right of the 
$(|\Delta S|_M,t)$ plane, due to the rising $S_l$, which means an enhancement of critical temperature of the lower layer. The first peak preserves its location since $S_u$ does not change. But the height of the first peak decreases when the value of
$S_l$ increases.

The effect of the exchange interaction between two layers of the system can be seen in Fig. \ref{sek4} (b). The variation of the IMEC with the temperature is depicted for $(1,7/2)$ bilayer for chosen values of $r$. As seen in 
Fig. \ref{sek4} (b), rising $r$ causes the first peak to slightly move in IMEC to the right of the $(|\Delta S|_M,t)$ plane. Remember that this behavior is valid for the bilayer that has layers with the same valued spins, as explained above. This means
that rising interaction between the layers destroys the effect that comes from the layer with a lower value of the spin. Although it is depressed, the first peak is survived in this $(1,7/2)$ bilayer system.

\begin{figure}[h]\begin{center}
\epsfig{file=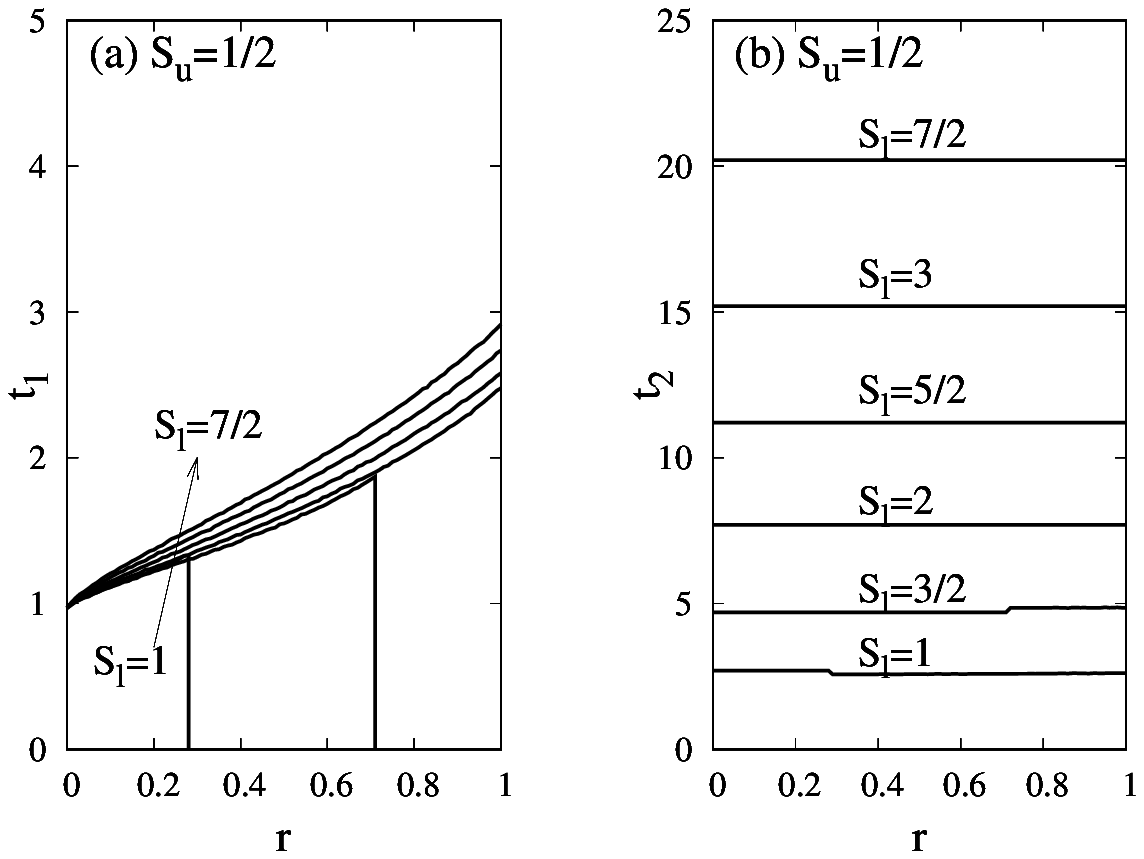, width=8.0cm}
\epsfig{file=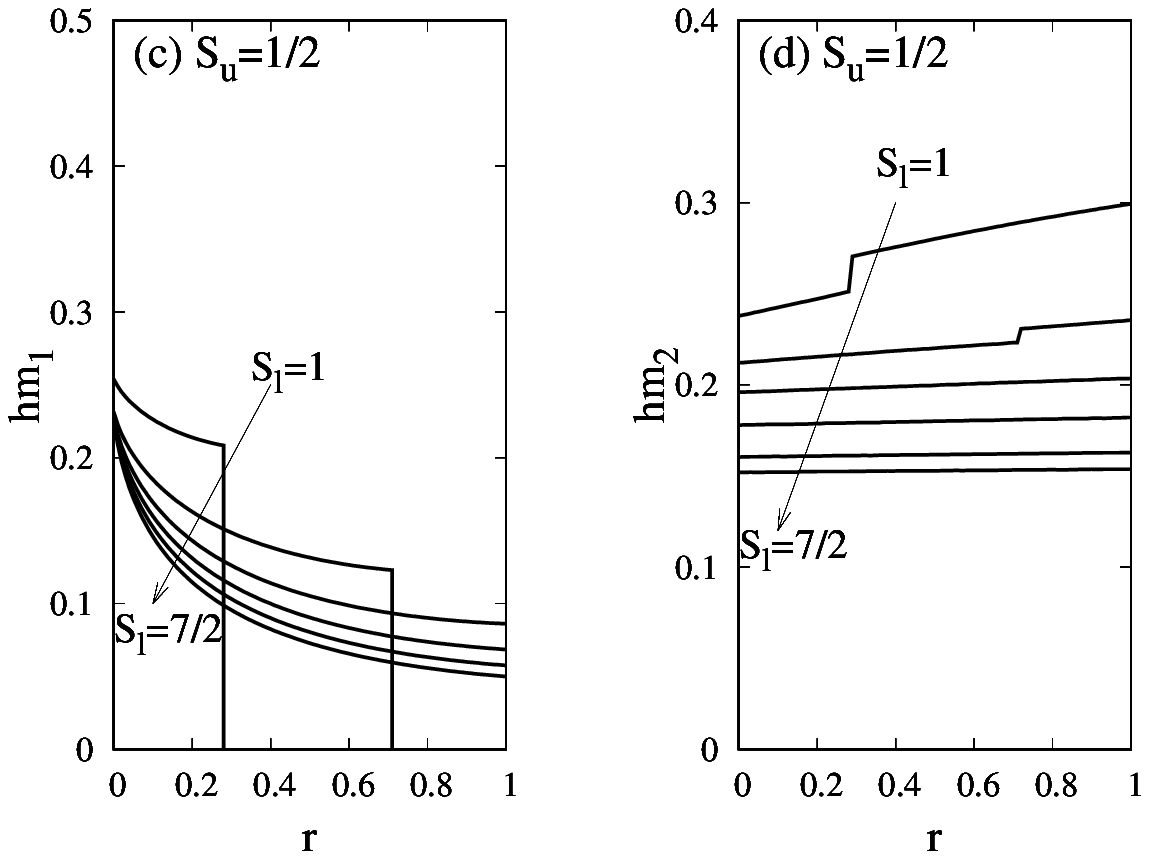, width=8.0cm}
\end{center}
\caption{The variation of the locations and heights
of the peaks with the $r$. $S_u=1/2$ and 
$S_l=1,3/2,2,5/2,3,7/2$ are chosen 
for the bilayer systems.} \label{sek5}
\end{figure}

\begin{figure}[h]\begin{center}
\epsfig{file=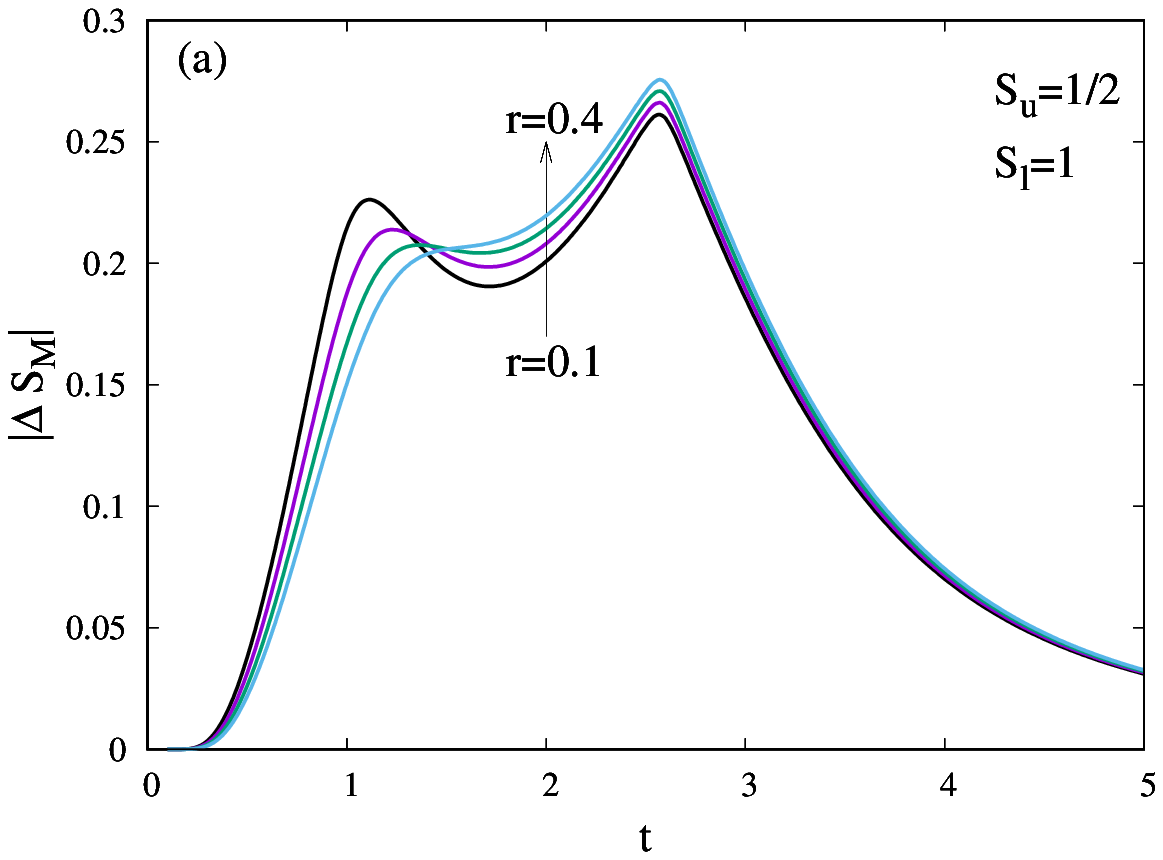, width=6.0cm}
\epsfig{file=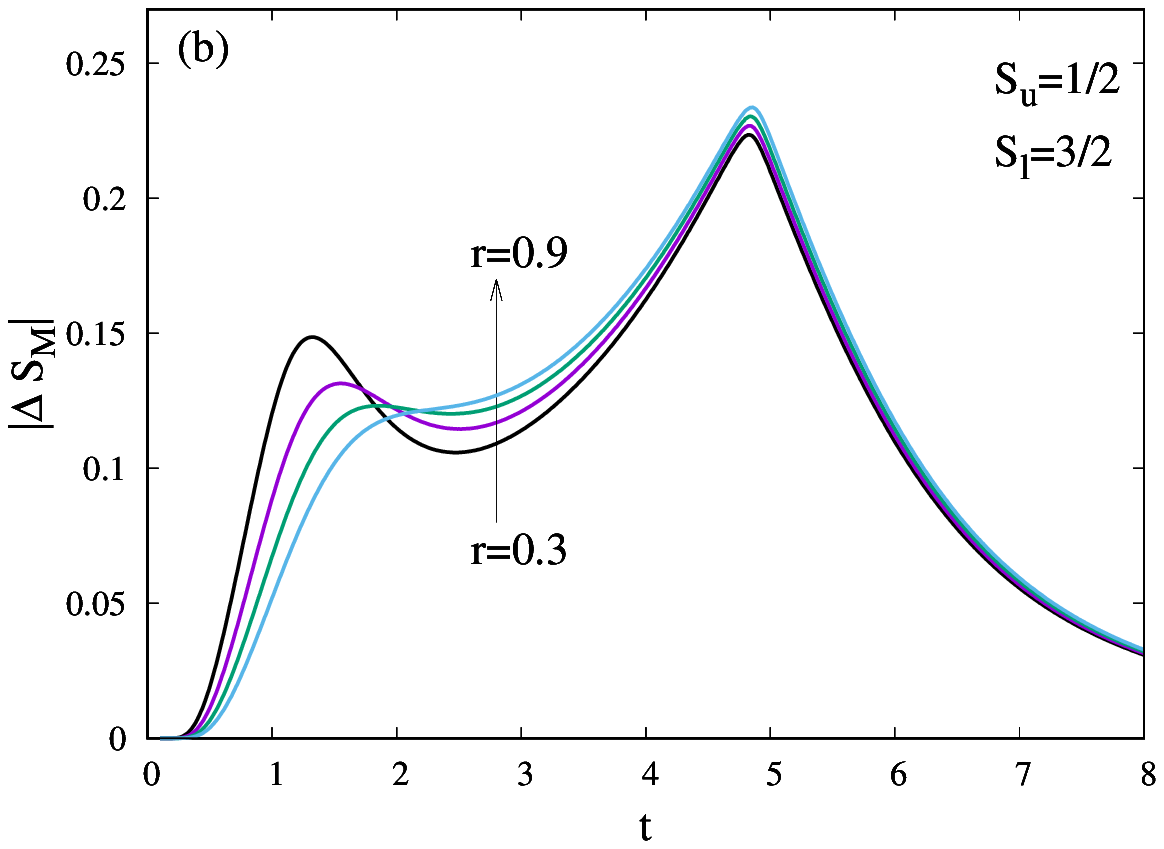, width=6.0cm}
\end{center}
\caption{The variation of IMEC with the temperature for selected values of (a) $S_u=1/2,S_l=1$ and the values of $r$ starting from $r=0.1$ with an increment by  $0.1$,  
(b) $S_u=1/2,S_l=3/2$ and the values of $r$ starting from $r=0.3$ with an increment by  $0.2$, 
for the bilayer systems.} \label{sek6}
\end{figure}

In order to get general outcomes regarding these two-peak behavior, we concentrate on the locations $(t_1,t_2)$
and the heights $(hm_1,hm_2)$ of these peaks. In Fig. \ref{sek5}, the variation of the locations and heights
of the peaks with $r$ can be seen for several bilayer systems. 
As seen in Figs. \ref{sek5} (a) and (c), when the interaction between the layers increases, the location of the first peak in IMEC shifts towards to the higher temperature values, and the height of the peak decreases. Besides, the second peak almost does not change, as can be seen in Figs. \ref{sek5} (b) and (d). These are consistent with the behavior depicted in Fig. \ref{sek4} (b). But, as we can see from Fig. \ref{sek5} (a) and (c) for $(1/2,1)$ and $(1/2,3/2)$  bilayers, the first peak in IMEC disappeared in specific values of $r$ (see the related curves in Figs. \ref{sek5} (a) and (c)). 
The evolution of the IMEC curves related to this situation can be seen in Fig. \ref{sek6}. As seen in  Fig. \ref{sek6}, when the interaction between bilayers rises, the first peak in IMEC becomes depressed, and after then, the peak becomes a plateau. This yields a disappearance of the first peak in IMEC.

\begin{figure}[h]\begin{center}
\epsfig{file=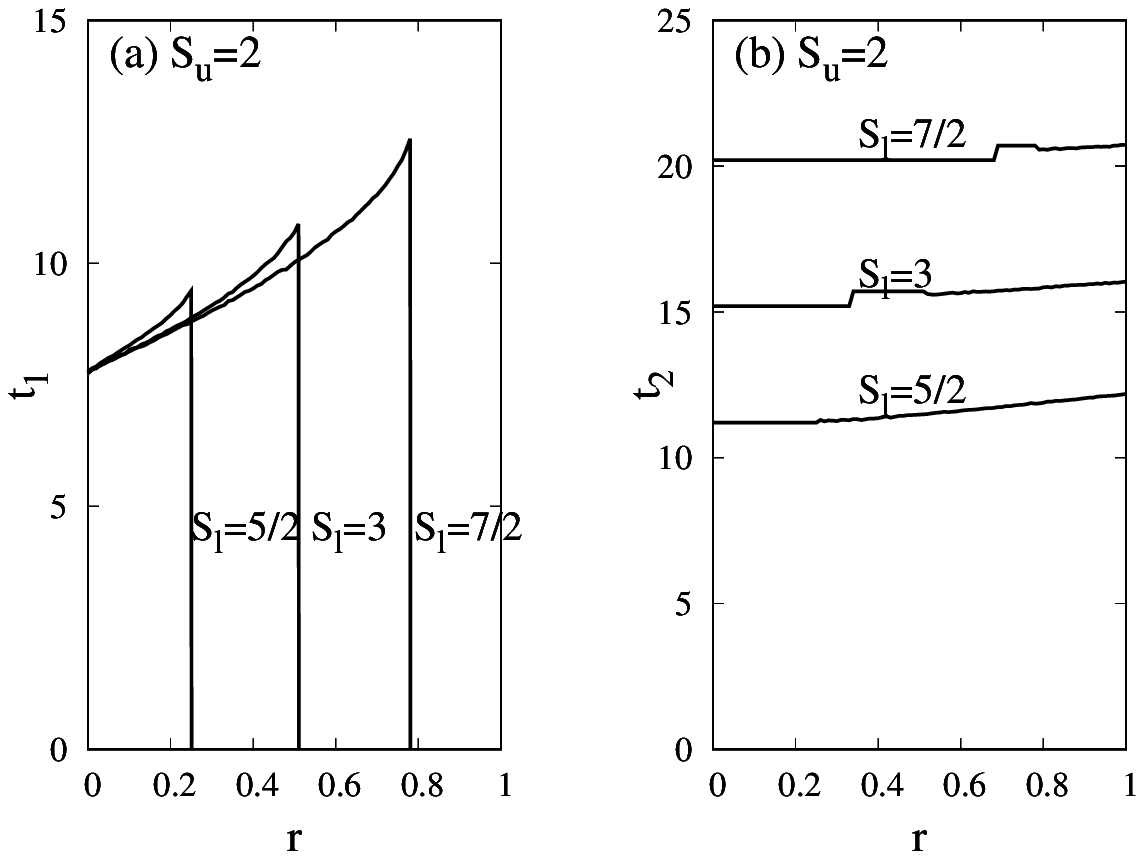, width=8.0cm}
\epsfig{file=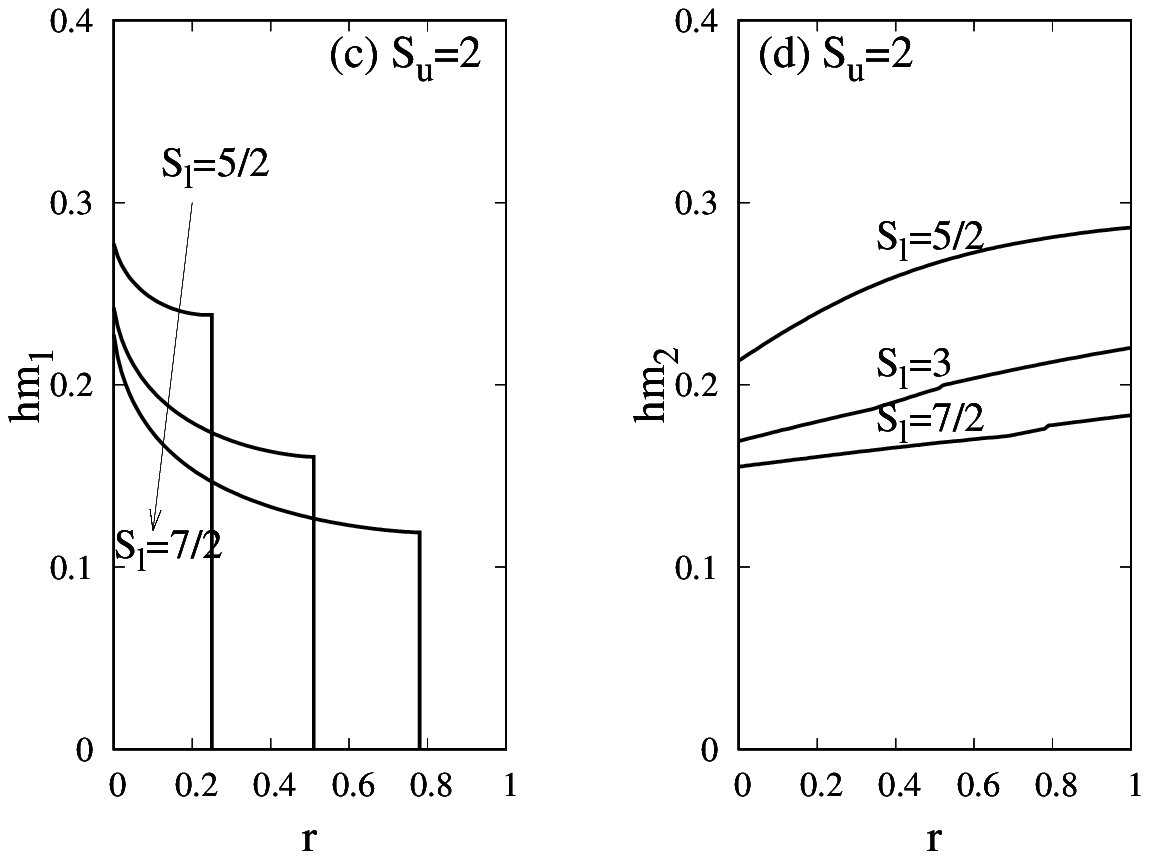, width=8.0cm}
\end{center}
\caption{The variation of the locations and heights
of the peaks with $r$. $S_u=2$ and 
$S_l=5/2,3,7/2$ are chosen 
for the bilayer systems.} \label{sek7}
\end{figure}

The same evolution can be seen in Figs. \ref{sek7} (a)-(d) for bilayers $(2,5/2), (2,3)$, and $(2,7/2)$. Again, the first peak disappears at specific values of $r$. When the value of $r$ increases, the magnetic properties of the bilayer is dominated by the layer which has higher spin value. While the exchange interaction becomes enhanced, the system evolves into one peak behavior in the variation of IMEC with the temperature. Indeed this characteristic behavior is valid for all spin pairs except $S_u=S_l$. The value that realizes the transition from the two peak behavior to the single peak behavior in IMEC is spin-dependent. These threshold values can be seen in Table \ref{tbl_1}.

\begin{table}
\begin{center}
\begin{tabular}{ c|c|c| c|c|c|c|c} 
 $S_u  \backslash  S_l$ & 1 & 3/2 & 2 & 5/2 & 3 & 7/2\\ 
 \hline
 1/2 & 0.283 &0.720& 1.074 & 1.249&1.411&1.527\\ 
  \hline
 1 & & 0.316&0.724& 1.061 &1.359 &1.611\\ 
  \hline
 3/2 &  & & 0.295 & 0.614& 0.924&1.205\\ 
  \hline
2 &  & &  &0.253 & 0.516& 0.783\\
\hline
5/2 &  & &  & & 0.229&0.446\\
\hline
 3 &  & &  & & &0.203\\

\end{tabular}
\end{center}\label{tbl_1}\caption{The threshold values that realizes the transition from the two peak behavior to the single peak behavior in IMEC. 
}
\end{table}

As seen in this Table, higher the spin magnitude means lower values of $r$ to destroy the double peak behavior in IMEC. Whereas, if one layer of the system consists of lower spin value, and the other has higher spin value, then greater values of $r$ are necessary to destroy the double peak behavior.



\section{Conclusion}\label{conclusion}

The IMEC has been determined for the bilayer system that consists of $S_u$ and $S_l$ valued spins in each layer. The dependency of IMEC on the spin value as well as on the value of exchange interaction between two layers has been investigated, and some general results have been obtained. 

It is now a well-known fact that magnetic materials exhibit maximum IMEC at the critical temperature. This shows itself in the variation of the IMEC with the temperature curve as a sharp peak. When the bilayer consists of one type of spin (i.e., $S_u=S_l$) it has been shown that rising spin value results in decreasing behavior in the peak of IMEC. This is the same behavior in bulk materials. On the other hand, rising exchange interaction value (between the layers) again results in the decline of the peak value. 

Significant results related to the bilayers that have  $S_u\neq S_l$ have been obtained. Double peak behavior has been obtained for these bilayers. This double peak behavior in IMEC has also been shown very recently by an experiment for manganite bilayers \cite{ref6}. The effect of the spin value and exchange interaction between the layers on the heights and the locations (temperatures)  of these peaks have been determined. It has been shown that when the exchange interaction between the layers increases, this double peak behavior causes a single peak behavior in IMEC. The threshold values of the exchange interactions, which give rise to this transition have been determined for several spin values of the layers. 

Indeed this double peak behavior and transition from this behavior to the single peak behavior are very important in terms of the technological applications. It is a well-known fact, near the IMEC peak, large magnetocaloric effect is obtained. Double peak behavior allows working on two different temperatures in terms of the magnetocaloric effect. On the other hand, turning this behavior into single peak behavior yields spreading the peak region of IMEC to a wider temperature range. Thus large refrigerant capacity is obtained. Although the fabrication of the bilayer from arbitrary spin valued planes and adjusting the exchange interaction between these planes are very hard tasks from the experimental point of view, it is important to obtain conclusions about the relation between the IMEC with these parameters. 

We hope that the results  obtained in this work may be beneficial form both 
theoretical and experimental points of view.


\newpage

 \begin{thebibliography}{00}

\bibitem{ref1} E. Warburg, Ann. Phys. 13, 141 (1881).
\bibitem{ref2} P. Debye, Ann. Phys. 81, 1154 (1926).
\bibitem{ref3} W. F. Giauque, J. Amer. Chem. Soc. 49, 1864 (1927).
\bibitem{ref4} A.M. Tishin, Y.I. Spichkin, The Magnetocaloric Effect and Its Applications, Institute of Physics, 2003.
\bibitem{ref5} P. Lampen et al, Appl. Phys. Lett. 102, 062414 (2013).



\bibitem{ref6}  Ruihao Yuan et al J. Appl. Phys. 127, 154102 (2020) 

\bibitem{ref7}  Yu-E. Yang et al, Journal of Physics and Chemistry of Solids 115 (2018) 311-316  


\bibitem{ref8}  A.A. Amirov et al, Journal of Magnetism and Magnetic Materials 470 (2019) 77-80
\bibitem{ref9}  Sushmita Dey et al, J Mater Sci (2019) 54:11292-11303
\bibitem{ref10}  Ali Osman Aya\c{h} et al (2017) , Philosophical Magazine, 97:9,671-682
\bibitem{ref11}  T. Kaneyoshi Journal of Physics and Chemistry of Solids 119 (2018) 202-209
\bibitem{ref12} Wei Wang, Feng-li Xue, Ming-ze Wang Physica B 515 (2017) 104-111 
\bibitem{ref13} Kaneyoshi, T. Journal of Superconductivity and Novel Magnetism 31, 2149 - 2155 (2018) 
\bibitem{ref14} A. Jabar et al Chemical Physics Letters 670 (2017) 16-21 
\bibitem{ref15}  Karimou, M. et al Journal of Superconductivity and Novel Magnetism  32, 1769 - 1779 (2019) 
\bibitem{ref16}    Karol Szalowski, Tadeusz Balcerzak Thin Solid Films 534, (2013), 546-552
\bibitem{ref17}  Ping Xu, An Du Physica B 521 (2017) 134-140
\bibitem{ref18} \"Umit Ak\i nc\i, Journal of Magnetism and Magnetic Materials,386,60-68 (2015)  
\bibitem{ref19} T. Balcerzak, J. Magn. Magn. Mater. 246 (2002) 213.  

\bibitem{ref20} \"U. Ak\i nc\i, Y. Y\"uksel, E. Vatansever, Physics Letters A, 382 3238-3243 (2018)  
 \end{thebibliography}
\end{document}